\documentclass{iaus}
\usepackage{graphicx}

\title[The structure of the thin disk] 
{The  Milky Way thin disk structure as revealed by stars and young open clusters}

\author[G. Carraro]   
{Giovanni Carraro}

\affiliation{Alonso de Cordova 3107, 19001, Santiago de Chile, Chile\\ email: {\tt gcarraro@eso.org} \\[\affilskip]}

\pubyear{2014}
\volume{298}  
\pagerange{XX -- YY}
\jname{IAUS\,298 Setting the scence for Gaia and LAMOST}
\editors{S. Feltzing, G. Zhao, N.\,A. Walton \& P.\,A. Whitelock, eds.}
\begin{document}

\maketitle

\begin{abstract}

In this contribution I shall focus on the structure of the Galactic thin disk. The evolution of the thin disk and
its chemical properties have been discussed in detail by T. Bensby's  contribution in conjunction with the properties
of the Galactic thick disk, and by L.Olivia in conjunction with the properties of the Galactic bulge.
I will review and discuss the status of our understanding of three major topics, which have been the subject of intense
research nowadays, after long years of silence: (1) the spiral structure
of the Milky Way, (2) the size of the Galactic disk, and  (3) the nature of the Local arm (Orion spur), where the Sun is immersed.  
The provisional  conclusions of this discussion are that : (1)  we still have quite a poor knowledge
of the Milky Way spiral structure, and the main dis-agreements among various tracers are still to be settled; (2) the Galactic disk 
does clearly \textit{not} have an obvious luminous  cut-off  at about 14 kpc from the Galactic center, and next generation Galactic models need to be updated in this respect, and 
(3) the Local arm is most probably an inter-arm structure, similar to what we see in several external spirals, like M~74.
Finally, the impact of GAIA and LAMOST in this field will be briefly  discussed as well.
\end{abstract}

\firstsection 
\section{Introduction}

A  quick glance  at the Hubble Atlas of Galaxies immediately reveals that spiral galaxies, when seen face-off, possess dusty and gaseous disks where
stars are barely visible. On the other hand, when seen face-on, they exhibit  quite  spectacular structures in the form of gaseous and stellar spiral arms,
bridges, inter-arm structures, knots, bifurcations, and so forth.
These detailed shapes are hardly repeated from one spiral galaxy to the other.\\

\noindent
Our Milky Way is believed to be a grand design spiral galaxy, of Hubble type Sb or Sc, and most probably very similar to NGC 1232, although
its precise structure has been challenging us for  more than 60 years, and it is still very far from being understood. The very special position of the Sun,
close to the Galactic plane, and immersed in a spiral feature (the Orion Spur), is one of the mayor difficulties astronomers have to face, together with the fact
that to probe the disk structure we have to search for features along the plane and penetrate across thick layers of gas and dust.\\

\begin{figure}[b]
\begin{center}
 \includegraphics[width=10cm]{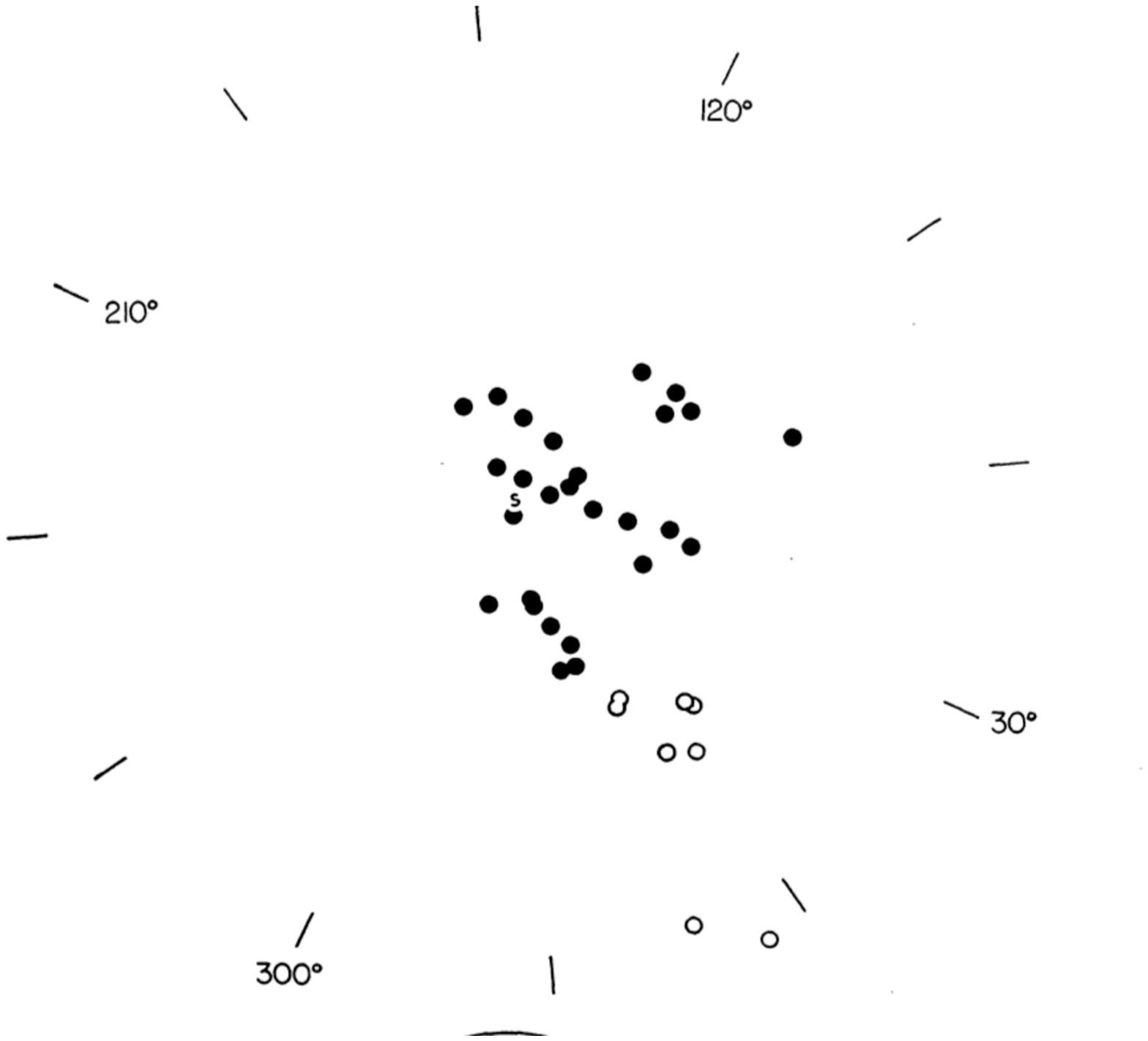} 
 \caption{The original view of the Milky Way as traced by OB stars in the solar vicinity.  Numbers indicate the Galactic longitude.Notice three clear structures: the Orion spur, where the Sun (S)
 is in, the Carina Sagittarius arm toward the Galactic Center, and the Perseus arm in the second quadrant,  toward the anti-center. From Morgan et al. (1953).}
   \label{fig1} 
\end{center}
\end{figure}

\begin{figure}[b]
\begin{center}
 \includegraphics[width=12cm]{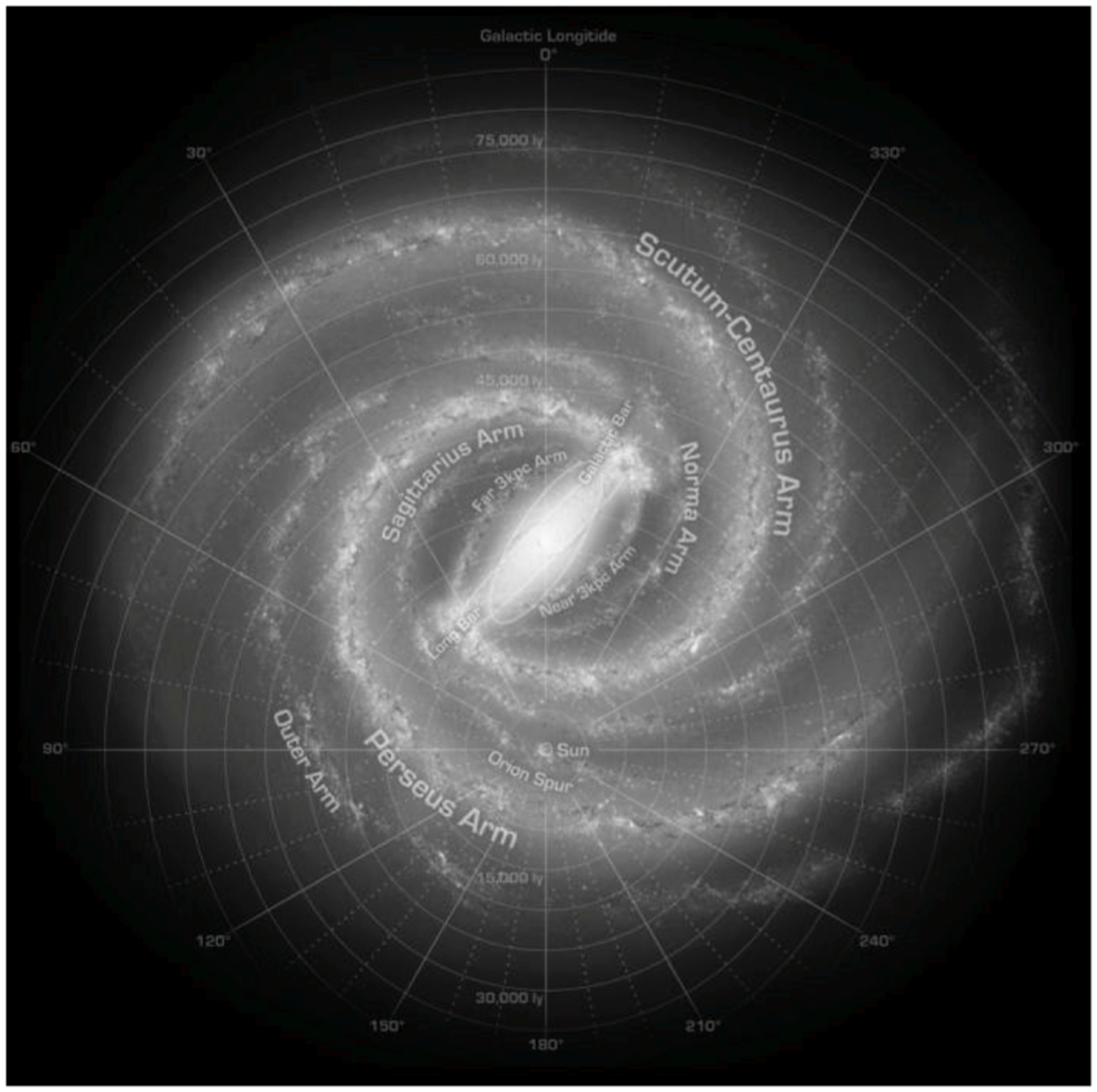} 
 \caption{The GLIMPSE artistic-rendered view of the Milky Way spiral structure (Churchwell et al. 2009).}
   \label{fig2} 
\end{center}
\end{figure}

\noindent
The thin disk of the Milky Way contains in fact mostly gas, dust,  and stars. It is the place where star formation occurs, inside the high-density spiral arms. 
It is most  probably detached from the Galactic thick disk and bulge (Gilmore et al. 1989), although there are 
different opinions on the subject (see the chapters  by Bensby, Origlia, Lepine, and Rix in this volume).\\

\noindent
Young stars (of spectral type O to A), either in clusters or in the general field, and gas (either atomic or molecular), are routinely used as tracers of its structure.
In Table~1,  I summarize the approximate distribution in mass of the typical thin disk components from Kalberla \& Kerp (2011), and references therein.
In the mass budget, stars are the largest contributors. At odds with the Galactic thick disk, stars in the thin disk have a large spread in age,
from virtually 0 to about 10 Gyrs (see Bensby review). Stars as old as 10 Gyrs are no longer tracers of the Galactic thin disk structure  as we observe it now, since they departed
significantly from their birth-place due to a variety of dynamical processes (migration, spiral arm perturbations, disk crossing, encounters with molecular clouds, 
 and so forth).
The present-day structure of the thin disk is therefore better described by young stars and gas. These two tracers are confined in a thin layer, less than 100 parsec
thick. Young stars, of OB spectral type,  are typically clumped, since they mostly form in associations and star clusters. These latter have a typical
life-time of a  few hundred million years, and afterwards they dissolve into the general Galactic field (de la Fuente Marcos et al.  2013). Older clusters are found, especially in the outer
disk (Carraro et al. 2013, and references therein), or at larger distances from the plane, where the environment is more favorable, and less encounters with molecular clouds are probable.\\

\noindent
In this review I will focus on young stars and stellar clusters, and how they trace the actual structure of the Milky Way. The perspective is purely
observational.  I will refer from time to time also to HI and CO surveys, without however entering into much detail. I am not going to mention much
bout maser surveys either (Reid et al. 2009). This is quite a promising but still young technique, which surely will significantly impact  our understanding of the disk in the future , once
more data will be accumulated.

\begin{table}
  \begin{center}
  \caption{Approximate distribution in mass of various thin disk components, taken from Kalberla  \&Kerp (2011), and references therein.}
  \label{tab1}  
 {\scriptsize
  \begin{tabular}{lcccc}\hline 
Component &  Mass ($10^9 M_{\odot}$)\\
\hline\hline
HI           &  8.0\\
HII          &  2.0\\
CO,CS  &  2.5\\
stars      &  70.5\\
\hline
\hline
  \end{tabular}
  }
 \end{center}
\end{table}

\begin{figure}[b]
\begin{center}
\includegraphics[height=6cm]{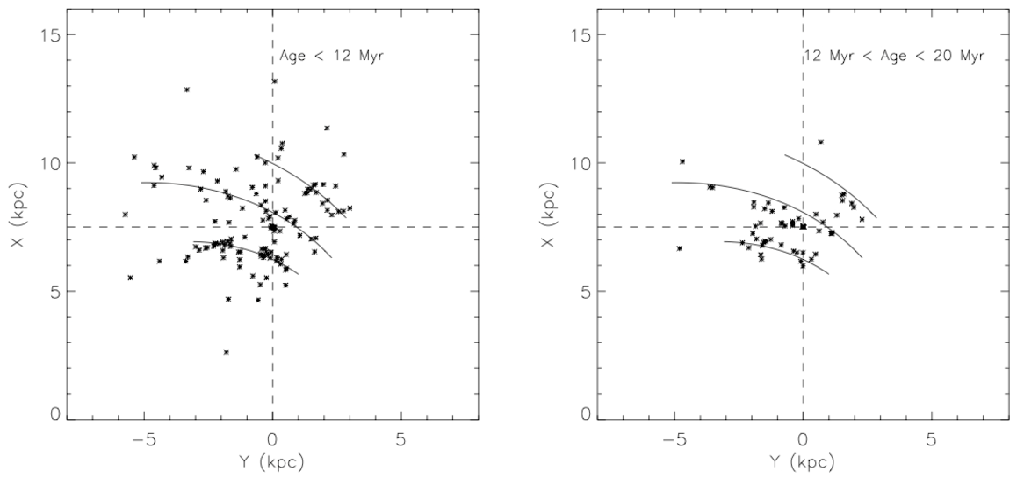}
 \caption{Spatial distribution of open clusters in the plane of the Milky Way as a function of their age. From Dias \& Lepine (2005).}
   \label{fig3} 
   \end{center}
\end{figure}

\begin{figure}[b]
\begin{center}
\includegraphics[height=10cm]{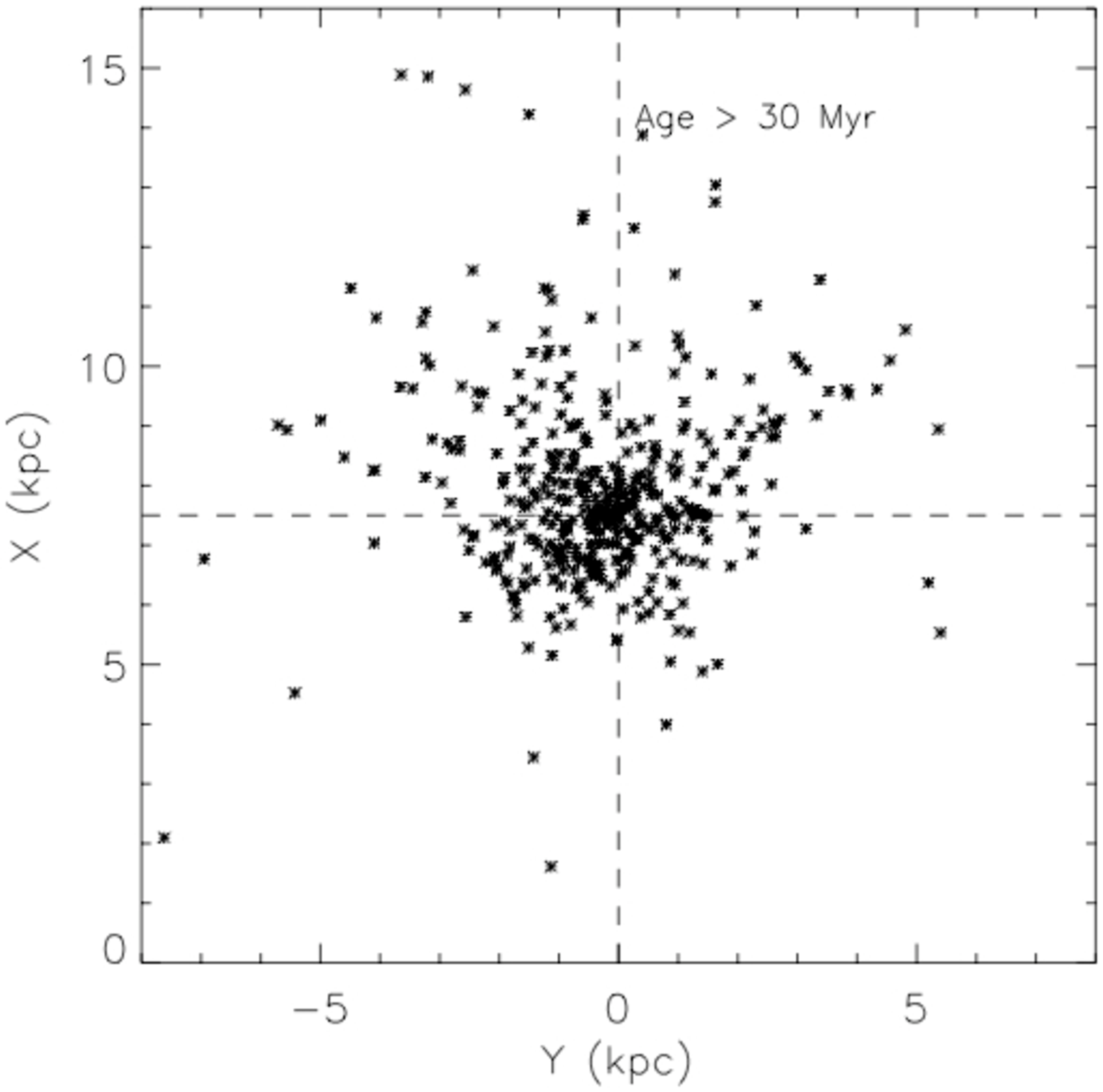}
 \caption{Spatial distribution of open clusters in the plane of the Milky Way as a function of their age. From Dias \& Lepine (2005).}
   \label{fig3} 
   \end{center}
\end{figure}

\section{The spiral structure of the Milky Way}

\noindent
The quest for the spiral structure of the Milky Way started enthusiastically in the early fifties, almost 70 years ago.  Interestingly enough,
for the perspective of this talk, the quest started using spectrophotometric distances to OB stars, since astronomers at the time realized that these stars
are present in the spiral arms of external galaxies (Morgan et al. 1952).  
The original picture of the Milky Way spiral structure  (see Fig.~1) only contains the local (Orion) arm (spur), where the Sun is located,
the Perseus arm in the second Galactic quadrant, and the Great Carina star forming region in the fourth Galactic quadrant.\\

\noindent
The HI 21cm line was  discovered later (for details, see the beautiful review from Gingerich 1985), but its success was so huge that it replaced the optical tracers  almost completely in the quest.
The HI view of the Milky Way spiral structure, developed after 50 years of research, is the one of a Galaxy with two mayor arms, the Perseus and the Scutum-Centaurus arms.
This is summarized by the artistic  rendering in Fig.~2, which results from  star counts using Glimpse data (Churchwell et al 2009), filtered by HI data. 
The two major arms are clearly indicated,
while arms like Carina or the outer (Norma Cygnus) arm appear as minor, secondary structures. 
Overall, the Galaxy appear as a beautiful, ordered,  grand design
spiral galaxy, of a kind that we very rarely see, e.g., by inspecting the Hubble Atlas.  Another recent, but strikingly different, HI realization of the spiral structure of the Milky Way is in Levine et al. (2006). \\

\noindent
Traditionally, this picture  is in net contrast with the optical/HII view of the Galaxy, that postulates that the Milky Way is a four arms
spiral galaxy. This constitutes the Georgelin \& Georgelin legacy, summarized in Russeil (2003), recently refined in much more detail 
-but just in the first Galactic quadrant -  by Anderson et al. (2011).\\

\noindent
According to Churchwell et al. (2009), the spectacular view of the Milky Way  in Fig.~2 comes from counting red clump stars. These are typically 200-300 
Myr or even older stars 
and, technically speaking, are not the ideal tracers of the young, gas-rich, star forming regions  which usually define spiral features. 
One can speculate that whatever stellar tracers can be in fact used, since stars tend to be trapped anyway in the high density potential wells generated by spiral
perturbations (see also the discussion at the end of this document).
Our opinion is that this is a weak argument.\\
To illustrate it, we refer to the results from Dias \& Lepine (2005)  in Figs.~2 and 3. Open clusters in the solar volume for different
ages are considered. From the left to the right one can readily appreciate how very young open clusters position in a seemingly regular spiral structure, while
at increasing clusters' age, the spatial distribution get more scattered in a way that no structure can be detected, since these clusters had a lot of time to move
away from their birth-places (see Fig.~3). \\

\begin{figure}[b]
\begin{center}
 \includegraphics[height=16cm]{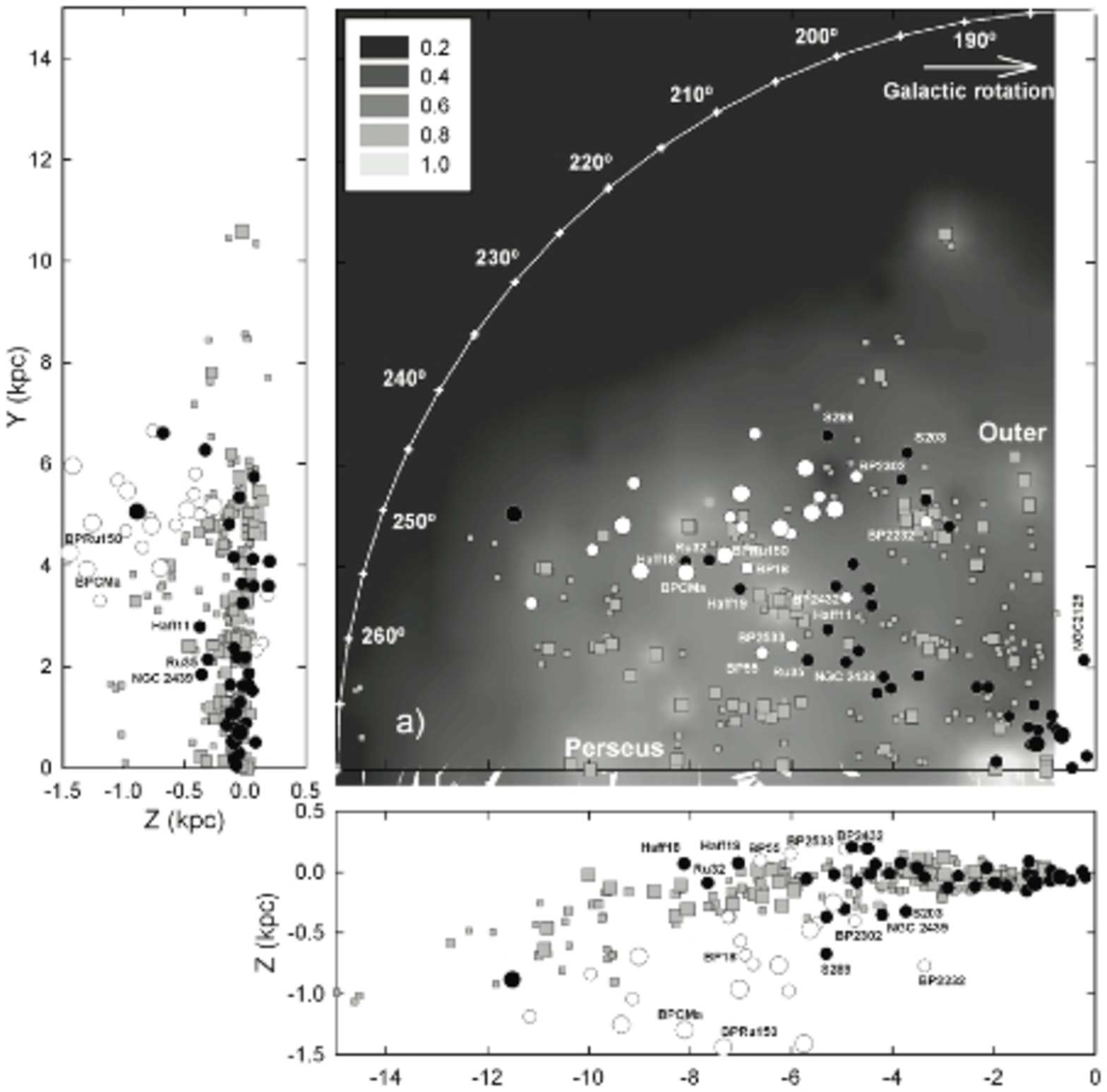} 
 \caption{The spiral structure of the Galactic disk in the third quadrant from V\'azquez et al.  2008. Solid dots indicate open clusters, while white dots
 indicate associations of young stars in the background of open star clusters. Finally, squares are for CO clouds. Notice the clear indication of the thin disk warp in the bottom panel.}
   \label{fig5} 
\end{center}
\end{figure}

\noindent
In recent years, the quest for spiral structure of the Milky Way using star and star clusters experienced quite a significant burst of activity.
This is mostly because of the failure of HI studies, that cannot  go much  beyond the detection of  gas density peaks in the velocity space (McClure-Griffiths et al. 2004; Dame et al 2011), and struggle in
 the process of translating these velocities into distances (Liszt 1985).\\
 
\noindent
Young open clusters and OB stars are powerful spiral arm tracers, and can probe spiral features in very remote regions of the Milky Way. Of course, deriving
their distance is not an easy task (Carraro 2011), but the technique is well established and suffers from less systematics than, for instance,  red clump stars (see below).
Deriving the distance to a young  (less than 100 Myr) open cluster/ association is straightforward in a sense. 
Star clusters and OB associations  are groups of coeval, co-spatial stars, and their distances can be robustly estimated in a statistical way. When UBV photometry is combined with spectral classification, the extinction law toward a star cluster can be derived and, in turn, its distance measured (see, as an illustration,  Carraro et al 2013, where the distance to the young open cluster Westerlund~2 is derived.).
The systematic use of star clusters to map the spiral structure of the Milky Way has been limited so far only to the third Galactic quadrant ($180^o \leq l \leq 270^o$).
In this sector of the Milky Way extinction is small (Moitinho 2001), and star clusters can be detected to very distant regions (Carraro et al. 2010), because the young disk is significantly warped (Moitinho et al 2006).\\

\noindent
A mayor break-through  is reported in V\'azquez et al. (2008), where for the very first time, distances to  a large sample of young (less than 100 Myr) open clusters (mostly from Moitinho 2001) are compared with  CO clumps all the way to 20 kpc from the Galactic center in the anti-center direction. Coupling homogenous  photometry with a solid technique, this study unraveled the spiral structure in the third quadrant (see Fig.~5), showing that the Perseus arm
does not seem to continue in this quadrant. Its structure is broken by the local arm (see below), which extends in the third quadrant all the way to the outer, Norma Cygnus arm. 
This is in contrast with the GLIMPE realization of the third quadrant, but in nice agreement, e.g.,  with the HI picture of the Milky Way from Levine et al. (2006).
In fact, this study finds that the strongest arms in Hydrogen are not Perseus or Scutum, but possibly Norma-Cygnus and Carina-Sagittarius, and confirms
the nature of the local arm.\\

\noindent
Attempts to repeat this kind of analysis are undergoing in the second quadrant (Mongui\'o et al 2012), using Stromgren 
\textit{ubvy} photometry. This photometric system has the obvious advantage that
stellar parameters can be measured precisely, but it  is distance-limited  due to the difficult to go deep enough with narrow band photometry. Besides,
the dusty Perseus arm in the second Galactic quadrant prevents reaching much further away, at least in optical.  \\

\noindent
The general, well-known optical picture of the Milky Way spiral structure in the solar vicinity is confirmed also 
using Cepheid stars, for which distances can be measured with high precision (Majaess et al. 2009).

\begin{figure}[b]
\begin{center}
 \includegraphics[height=12cm, angle=-90]{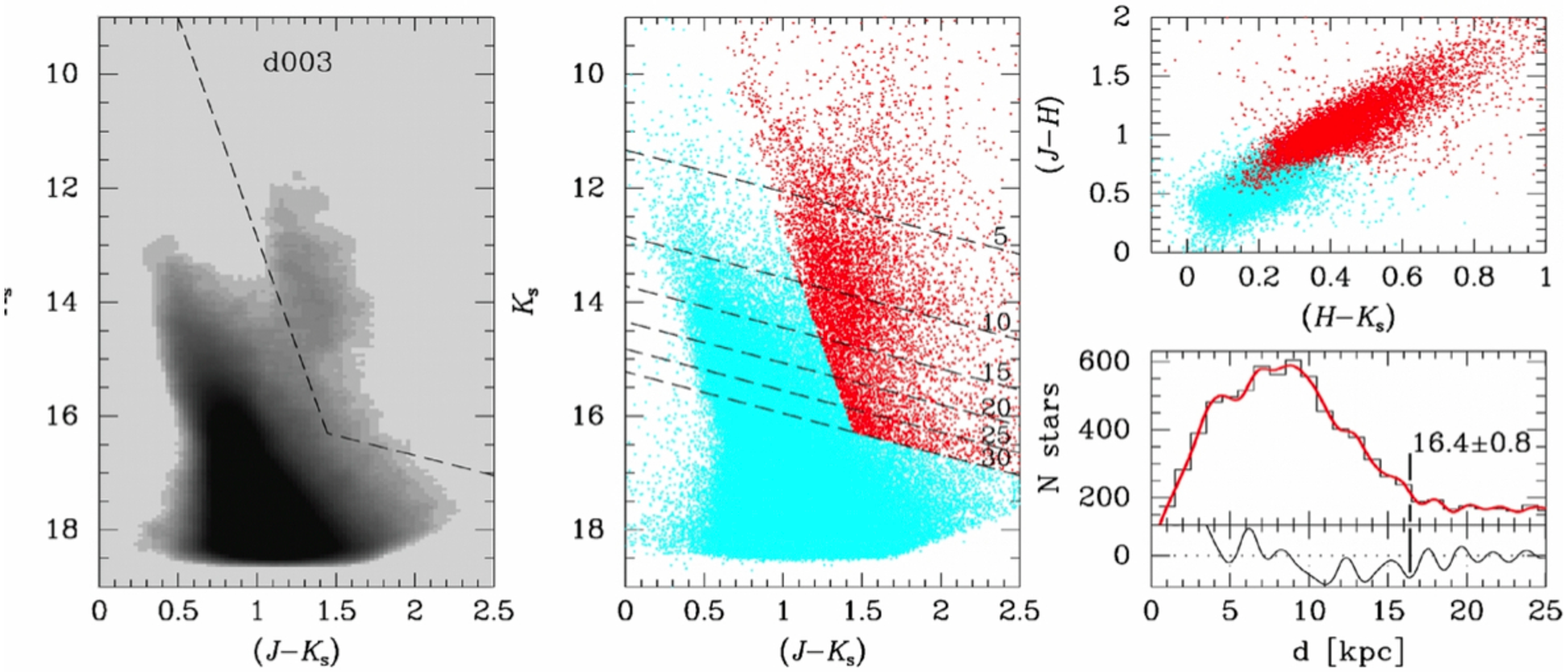} 
 \caption{ Clump stars' selection, and distance estimates from Minniti et al. 2011. Note the color and magnitude cuts.}
   \label{fig6} 
\end{center}
\end{figure}

\begin{figure}[b]
\begin{center}
 \includegraphics[height=12cm, angle=-90]{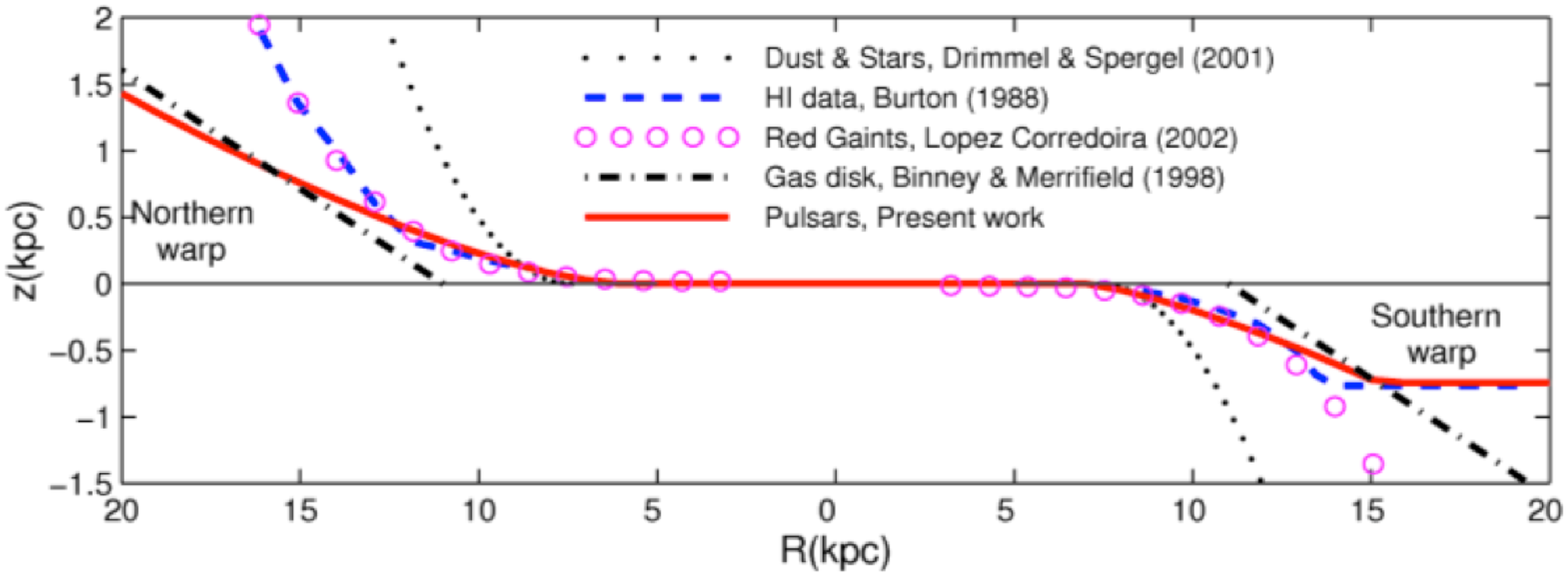} 
 \caption{The magnitude and extend of the Galactic warp in the northern and southern disk from various indications from Yusifov (2004). Note how
 observations limited to low Galactic latitudes (Z = 0 indicates the formal $b=o^o$ Galactic plane) naturally miss the disk and  find erroneously a cut-off. }
   \label{fig7} 
\end{center}
\end{figure}

\begin{figure}[b]
\begin{center}
\includegraphics[height=10cm,angle=-90]{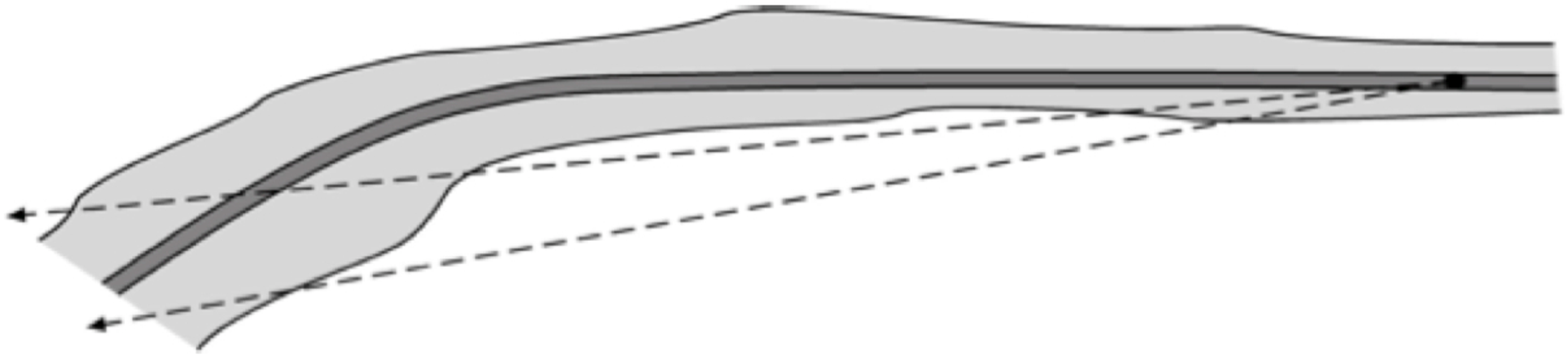} 
 \caption{An artistic rendering description of the warped Galactic disk as seen from $l = 270^o$. The Galactic center is to the right, the anti-center to the left. The dashed lines indicate the real lines of sight  from the Sun (the dot) toward the warped Galactic thin disk (the dark structure) and thick disk (the light structure). This figure is courtesy of Andre Moitinho.}
  \label{fig8} 
\end{center}
\end{figure}

\section{Does the Galactic disk have a density cut-off?}

\noindent
Models of Galactic structure and stellar population like the Besancon one (Robin et al. 2003) postulate that the Galactic disk has a strong density drop at about 14 kpc from the Galactic center, in the anti-center direction. Recently, Minniti et al. (2011) found that this cut-off is a general feature, and it is detectable in several other
Galactic directions (their Fig.~3). These results are clearly biased both in the underlying technique and in the assumptions. 
In the case of Minniti et al. (2011), two mayor biases are present.
First, the authors claim they take the disk warp into account, still their fields are limited to $-2\leq b \leq2$. The warp is, however,  much more extended in latitude (see Momany et al. 2006, and references therein).
Second, they use as tracers clump stars. These stars suffer form variable extinction when observed so close to the plane. Momany et al. (2006) exhaustively discussed
this point , and clearly showed  how Red Giant Branch stars are much better Galactic structure tracers. 
In fact, in the color magnitude diagram (see Fig.~6), red clump stars 
form an almost vertical strip, stretched by extinction, that at some point crosses the sub-giant branch stars in the disk.  Besides, looking across the disk,
a huge amount of variably reddened dwarf  stars are also intersected. 
This implies first that it is very hard to predict the contamination level of the sample, and, second,  that a sharp magnitude cut has to be adopted, 
to avoid confusion. This, in turn, naturally  implies a distance cut or limit. 
Stars more distant than the cut are simply ignored. The detected cut is therefore an artifact of the adopted sample.\\

\noindent
To better probe whether the disk has a cut off or not, one should look at the anti-center direction, where extinction is less important. This is what Robin et al.
(1992) did, sampling photometrically a field in the direction of the Galactic anti-center, but close to latitude $b \approx 0$.
The cut-off they found is in this case produced by the Galactic warp and flare, as illustrated in Figs. 7 and 8, and it is again artificial. 
It is natural to find a cut-off when looking along the formal $b=0^o$ plane, since the disk at  a distance of 13-15 kpc from the plane starts bending down. 
This is quite typical also in many external galaxies. \\

\noindent
To follow the disk continuation ones must follow the warp and flare, and therefore the cut-off is simply an \textit{illusion} caused by the warp and flare.
Recent works along this line (Carraro et al. 2010; Sale et al. 2010; Brand \& Wouterloot 2007) indeed find that the stellar disk is much more extended , up to 20 kpc
from the Galactic center, and in agreement
with HI and HII surveys. 
In particular,  Carraro et al. (2010) found extremely young star associations at more than 20 kpc along the warped disk in the third quadrant.\\

\noindent
This implies that the disk is active even at those distances, in agreement with the results of star forming region search 
in the outer disk of external spirals (Chung et al. 2009; Laine et al. 2013).\\

\noindent
A second, quite important conclusion, is that models of Galactic structure need to be revised by removing this artificial cut-off and properly model
the Galactic warp and flare. This will help prevent observations'
mis-interpretations, like the one of the Canis Major galaxy  and  the Monoceros ring (Moitinho et al. 2006, Momany et al. 2006, Hammersley \& L\'opez-Corredoira 2012, L\'opez-Corredoira et al. 2012).

\section{The nature of the Local arm (Orion Spur)}

\noindent
The Orion spur is the spiral feature inside which our Sun is located. The very inner location of the solar system makes it difficult for us to understand the real nature, orientation, and extent of this structure. 
It was originally detected by Morgan et al. (1953, see Fig.~1), but its nature has been elusive to present days. It might be a real arm, like Perseus,  or a
kind of inter-arm feature,
like the ones seen in many external spirals.
I would like to summarize here recent advances in the field, and propose a new picture of the spiral feature we are immersed in.
Recently,  Xu et al. (2012) studied a sample of about 30 masers located in the Orion arm, and measured their trigonometric parallaxes and kinematics.  Based
on that,
they conclude that the kinematic of the sample is  typical of a grand design spiral arm, and conclude that the Orion spur is an arm of the same nature of Perseus
or Scutum.
While the technique is promising, their sources, unfortunately,  cover quite a small volume of a few kpc around the Sun, making it difficult to understand whether
the whole Orion spur behave or not like a grand design arm.\\

\noindent
A much larger volume has been covered by Moitinho et al. (2006) and V\'azquez et al. (2008) using young open clusters and field stars
in their background. This study, as already outlined  before,
provides a fresh  picture of the spiral structure of the Milky Way in the third quadrant. 
In particular, it shows that the Local arm penetrates into  the third quadrant, and seems to break the Perseus
arm before reaching the outer, Norma Cygnus, arm. If this is confirmed, it implies that the local arm is a much larger structure than currently believed, and
behaves like a bridge, connecting the Norma Cygnus arm in the outer disk possibly all the way to  the Sagittarius arm in the inner first quadrant of the galaxy. This latter extension has
however been poorly explored so far, and much work is warmly recommended. This picture of the Orion spur is not an isolated one: the giant spiral M~74  bears an impressive resemblance to our Milky Way (see Fig~5  in V\'azquez et al. 2008).\\

\noindent
Intriguingly enough, this picture of the local arm as traced by young open clusters is very similar to the HI realization from Levine et al. (2006). Their Fig.~4 shows pretty clearly that a conspicuous  Hydrogen structure departs from roughly the Sun location and enters the third quadrant, breaking Perseus and reaching
the outer arm. \\

\noindent
It is therefore difficult to conceive that the Orion spur is a spiral arm like, e.g., Carina-Sagittarius or Scutum-Crux. As a by-product, 
this result lends support to  the idea that  the Perseus arm is  most probably \textit{not} a grand design spiral arm.

\section{Conclusions}

In this talk, I  reviewed the structure of the thin disk of the Milky Way as traced by young stars and young open clusters. This stellar population traces the disk not only close to the Sun, but all the way to the disk edge in the Galactic anti-center.  The expectations from LAMOST and GAIA are very high in this field. For many stars,  precise distances  will be available and it will be possible to position them in the plane of the Galaxy to confirm or deny the actual pieces of evidence
for the disk spiral structure and the disk shape and extent.\\

\noindent
Specifically, available data on young open cluster and young field stars indicated that the disk is not truncated at 14 kpc from the Galactic center, and that the spiral structure on the outer disk is much different from GLIMPSE expectations. The large-area anti-center coverage of the LAMOST survey will surely provide additional material to improve our understanding on the outer disk.\\

\noindent
Besides, the kinematics of millions of stars in the solar vicinity from the GAIAS mission will allow one to put firm constraint on the local arm nature and motion,
and, hopefully, to understand ultimately its origin.\\

\noindent
In the meantime, it is highly recommended to keep accumulating good data on as many young open clusters as possible, for which distances can still be determined precisely.
The knowledge of the disk structure is particularly intriguing toward the Galactic center, where - in stars- no clear evidences of spiral structure have been found
so far (Perren et al. 2012). Young  OB  stars  seem to show a continuum
in their distribution all the way to the Galactic center, with no significant peaks beyond the Carina Sagittarius arm at about 2 kpc from the Sun.

\section{Acknowledgements}
I am mostly indebted to my close collaborators Andre Moitinho and Ruben A. V\'azquez. I also benefited from very interesting discussions
with Yazan Momany, David Turner,  Martin L\'opez-Corredoira, and Butler Burton.

\begin{discussion}

\discuss{Antoja}{Some studies suggest that the massive spiral structure traced by \textit{old} stars induces two additional spiral arms made of gas and stars.
Do you think we have enough evidence for that? What do you think of that idea?}

\discuss{Carraro}{I believe spiral arms regenerate continuously. In the case of the Milky Way we do not see differences in young star and star clusters among
the so-called massive arms (Perseus and Scutum) and the so-called secondary arms (Carina and Norma).  The difference is apparently seen only in HI,  although different surveys produce different results.}

\discuss{Binney}{I think that N-body simulations make it clear that spiral structure is ephemeral. But it does not from this follow that it cannot be traced by old object
 because spiral structure is written in the gravitational potential. This marshall gas, causing it to shock and form stars, but also marshall stars of all types, causing them to linger along arms. So yes very young objects are good tracers of spiral arms via star formation, but older stars that have smallish velocity dispersion can also be good tracers. }
 
 \discuss{Carraro}{I think the argument of the gravitational trapping is mostly correct. 
 However to conclude that old stars trace spiral arms is  incorrect and dangerous in my point of view.
 Old stars are mostly spread everywhere, while young stars and gas are  preferentially located inside spiral arms, and therefore are the best spiral arm tracers.
 In the specific case of star counts (GLIMPSE) , there is no way to distinguish small velocity dispersion from high velocity dispersion clump stars, 
 one sums up everything, and therefore the outcome is not completely reliable. I would tend to believe more in the distribution of the younger material.}

\end{discussion}

\end{document}